\documentclass[runningheads,nynorsk]{llncs}
\usepackage{multicol}
\usepackage{rotating}
\usepackage{lmodern}
\usepackage{babel}
\usepackage[utf8]{inputenc}
\usepackage[T1]{fontenc}
\usepackage{url}
\usepackage{csquotes}
\usepackage{llncsnorsk}

\usepackage{tikz}
\usetikzlibrary{decorations.pathreplacing,calligraphy}

\usepackage{comment}
\usepackage{graphicx}
\usepackage{amsmath,amssymb}
\usepackage{subcaption}

\usepackage{url}
\usepackage[vskip=0pt,font=itshape]{quoting}

\usepackage[backend=biber, natbib=true, style=apa]{biblatex}

\def\andname{og}
\def\lastandname{ og}

\begin{document}

\title{Om å kartleggja mørk materie med maskinlæring}
\author{
      Hans Georg Schaathun\inst{1} \and
	   Ben David Normann\inst{2}\and
           Kenny Solevåg-Hoti\inst{3}
        }
\institute{%
\email{georg@schaathun.net}\quad
     \and\email{ben.d.normann@ntnu.no}\\
   Institutt for IKT og realfag \\ 
   Fakultet for informasjonsteknologi og Elektroteknikk\\[0.5ex] 
     \and
     \email{kenny.solevag-hoti@ntnu.no}\\
   Universitetsbiblioteket\\[0.8ex]
   NTNU --- Noregs Teknisk-Naturvitskaplege Universitet \\
   6025 Ålesund, Norway}

\maketitle


\def\DL{\ensuremath{D_{\mathrm{L}}}}
\def\DS{\ensuremath{D_{\mathrm{S}}}}
\def\xiV{\boldsymbol{\xi}}
\def\nuV{\boldsymbol{\nu}}
\def\etaV{\boldsymbol{\eta}}

\hyphenation{ray-trace}

\begin{abstract}
   Gravitasjonslinsing er fenomenet der ljos frå fjerne himmellegeme
   vert avbøygd av tyngdekraften frå andre himmellegeme, som ofte
   ikkje er fullt synlege fordi mykje av massen er mørk materie.
   Observert gjennom ei gravitasjonslinse, framstår fjerne galaksar
   som forvrengde.
   Der er mykje forskingsaktivitet som freistar å karleggja mørk materie
   ved å studera linseeffektar, men dei matematiske modellane er kompliserte
   og utrekningane krev i dag mykje manuelt arbeide som er svært tidkrevjande. 
   I denne artikkelen drøftar me korleis me kan kombinera rouletteformalismen
   åt Chris Clarkson med maskinlæring for automatisk, lokal estimering
   av linsepotentialet i sterke linser, og me presenterer eit rammeverk
   med programvare i open kjeldekode for å generera datasett og validera
   resultat.

\end{abstract}

\section{Innleiing}

Eit av dei store måla i kosmologien er å kartleggja universet.
Moderne teleskop gjev tilgang til eit enormt biletmateriale,
men om lag $85\%$ av massen er ikkje synleg.
Sokalla mørk materie gjev ikkje frå seg ljos.
Fysikarane veit lite om denne materien, men han må vera
der for at den generelle relativitetsteorien skal stemma med observasjonane.
For å lokalisera den mørke materien, må ein sjå korleis han påverkar
andre, synlege fenomen.  Sidan ljoset vert påverka av tyngdekraften,
vil mørk materie kunna danna sokalla gravitasjonslinser, som forvrenger
bilete av meir fjerne galaksar.  Dette er analogt til vanlege, optiske
linser.  Sjå t.d.\ \citet{bertone18}.

Ved å studera forvrengde bilete av fjerne galaksar, er det råd å utleida
plassering, form, og storleik på gravitasjonslinsene, men utrekningane
er komplekse og med dagens teknikkar tek det fleire dagar med manuelt arbeide
å rekonstruera éi linse.
Ei rekkje forfattarar har brukt maskinlæring på ulike måtar for å analysera
gravitasjonslinser.
\citet{hezaveh17} parameterbestemmer ein singulær, isoterm, elliptisk 
linsemodell (SIE) med sterk linsing.
\citet{biggio2023} bruker representasjonslæring for å finna eit nevralt nettverk
som tilnærmer ei spesifikk linse.  Der er også ei rekkje klassifiseringsmodellar
for å finna linser på nattehimmelen.

Kvar tilnærming byggjer på bestemte modelleringsparadigme med ulike 
styrkar og lyte; ingen gjev ei komplett og generisk løysing.
Ei særleg utfordring er sokalla klyngelinser, der fleire 
massekonsentrasjonar bidreg til linseeffekten på det same biletet.
I denne artikkelen føresleg me, heller enn å søkja ei global skildring
av linsa, å gje ei lokal skildring av \emph{effekten av} linsa i eitt
punkt på nattehimmelen.
Den lokale skildringa byggjer på rouletteformalismen etter
\citet{clarkson16a}, som søkjer å viska ut det konvensjonelle skiljet
mellom svak og sterk linsing.

Hensikta med artikkelen er å etablera eit rammeverk og programvare for
å støtta vidare arbeide med maskinlæring på rouletteformalismen, 
innanfor både informatikk og fysikk.
Kjernen er roulettesimulatoren \citep[sjå][]{hgs2023ecms}, som let
oss syntetisera treningsdata.
Her skal me imidlertid fokusera på prosessen som heilskap, og korleis
maskinlæringa kan brukast til å bestemma koeffisientane i rouletteformalismen,
og ikkje berre bestemma parameter i ein postulert linsemodell.
Avsnitt~\ref{sec:roulette} presenterer roulettemodellen og -simulatoren.
Avsnitt~\ref{sec:ml} gjev eit konseptprov for å visa at maskinlæring er lovande,
men i all hovudsak står dette att som eit ope problem.

\section{Modell og Problem}

\begin{figure}[bt]
   \begin{center}
\includegraphics[width=0.85\textwidth]{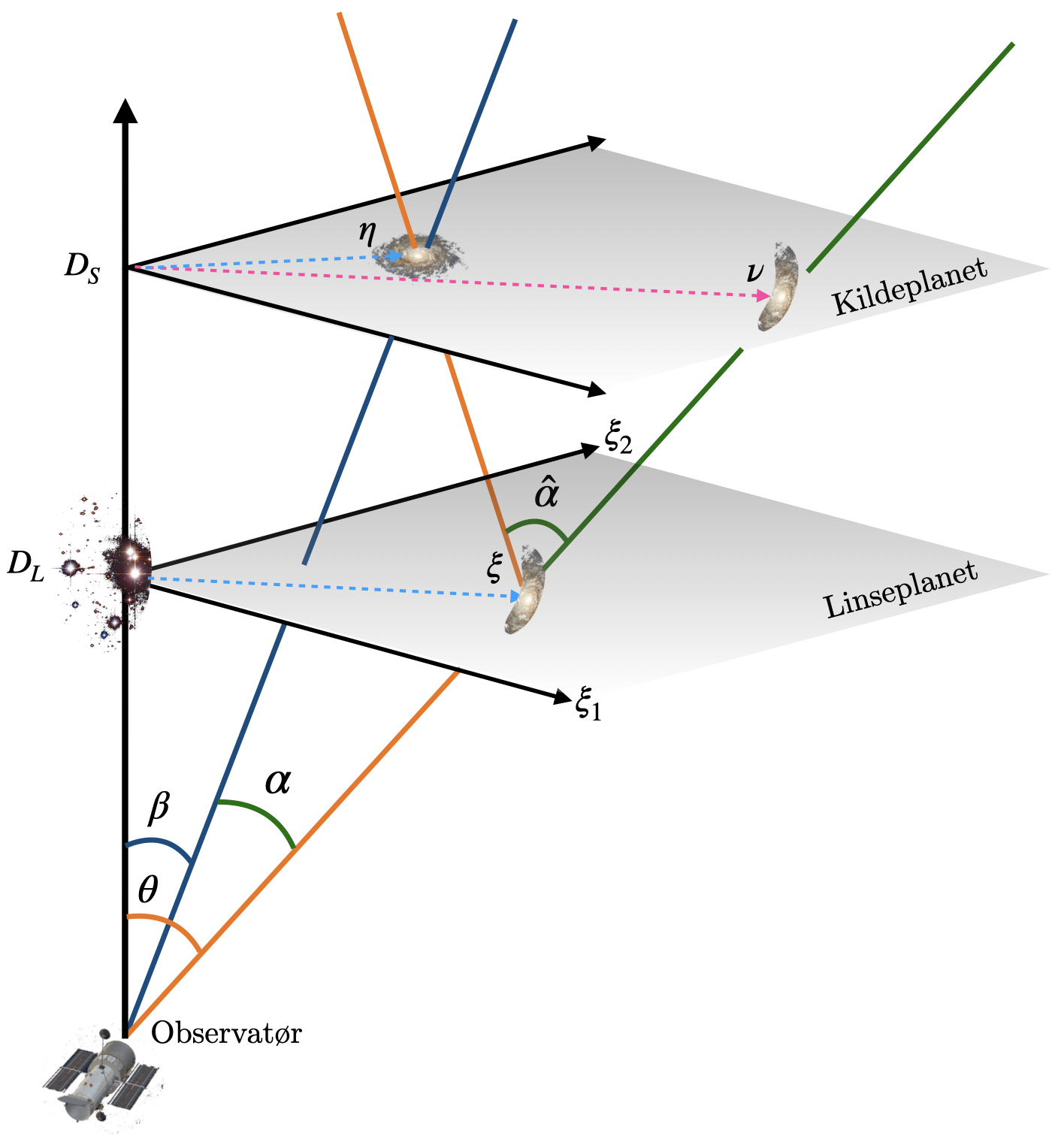}
   \end{center}
    \caption{Observasjon av nattehimmelen gjennom ei gravitasjonslinse.}
    \label{fig1}
\end{figure}

All materie, både ljos og mørk, oppfører seg som ei linse som forvrenger bilete 
av fjernare galaksar.
Eddington målte avbøyinga i solljoset under ei solformørking i 1919 og viste 
samsvar med den generelle relativitetsteorien etter Einstein.
Omfattande teoretisk arbeide er gjort sidan den gongen, og til tross for periodar
med pessimisme pga.\ skrinne observasjonar og dårleg oppløysing, er gravitasjonslinsing
no vorte eit av dei mest lovande verktya for informasjonsmining frå nattehimmelen.

Figur~\ref{fig1} viser ein enkel modell med éin observasjon og éi linse.
Massen er konsentrert i to plan, linseplanet som me søkjer å kartleggja og
kjeldeplanet som er opphav til den synlege observasjonen.
Me føreset at linsemassen er konsentrert i eit plan og ikkje har utstrekking
langs synslina.  Denne sokalla
tynnlinsetilnærminga er rimeleg fordi avstandane er astronomiske og tjukkna dermed
neglisjerbar.
Strengt teke er linse- og kjeldeplanet kuleoverflater, men når synsvinkelen er smal
kan me likevel føresetja flate plan (flathimmeltilnærminga).
I det fylgjande lèt me \DL\ og \DS\ stå for kortaste avstand til hhv.\ 
linseplanet og kjeldeplanet.

Linsa dannar eit linsepotensial som er skildra som ein funksjon
$\psi(\xiV)$ som gjev ein reell verdi for kvart punkt $\xiV$ i linseplanet.
I figur~\ref{fig1} ser me korleis observert ljos som kjem gjennom $\xiV$, 
ser ut til å koma frå $\nuV$ i kjeldeplanet, medan det i røynda kjem frå
$\etaV$ og vert bøygd av linsa.
Avbøyinga er bestemt av \emph{raytrace}-likninga som er gjeven som
\begin{align}
   \label{eq:raytrace}
   \etaV - \nuV =-\DS\DL\cdot
      \bigg(\frac{\partial\psi(\xi)}{\partial\xi_1},
      \frac{\partial\psi(\xi)}{\partial\xi_2}\bigg).
\end{align}
Sidan avbøyinga er forårsaka av massen, må der vera ein samanheng mellom 
massetettleiken $\kappa$ og linsepotensialet $\psi$.
Denne samanhengen skriv me som
\begin{align}
  \kappa(\xi)=\frac{\DL^2}{2}\nabla^2\psi(\xi)
=\frac{\DL^2}{2}\bigg(\frac{\partial^2\psi(\xi)}{\partial\xi_1^2}
+\frac{\partial^2\psi(\xi)}{\partial\xi_2^2}\bigg).
\end{align}

\begin{remark}\label{rem1}
   Oppmerksame lesarar kan sjå at faktoren $\DL$ er utelaten i programkoden,
   bortsett frå i forholdet $\chi=\DL/\DS$ som skalerer storleikar mellom
   linse- og kjeldeplanet.
   Dette er fordi $\DL$ og $\DS$ forøvrig vert kansellert i alle vidare
   utrekningar, noko som kan vera tidkrevjande men ganske rett fram å stadfesta.
\end{remark}

\begin{remark}
   Det er vanleg at linsa inneheld ljos materie i tillegg til mørk materie, 
   slik at det er mogleg å fastsetja $\DL$, og me vil gå ut frå at dette er
   tilfellet og at både $\DL$ og $\DS$ er kjende størrelsar.
\end{remark}

\begin{example}
   Den singulære isotermiske sfæremodellen (SIS) 
   \citep[sjå t.d.][Sec.\ 8.1.4]{bok:schneider92_SEF}
   har linsepotensiale
   \begin{align}
   \label{psiSIS}
   \psi_\textrm{SIS}(\xiV)=\frac{R_\textrm{E}}{D_\textrm{L}^2}|\xiV|,
      = \frac{R_\textrm{E}}{D_\textrm{L}^2}\sqrt{\xi_1^2+\xi_2^2},
   \end{align}
   der $R_E$ er einsteinradien som gjev styrken (eller totalmassen) på linsa.
   \emph{Raytrace}-likninga i SIS-modellen vert
   \[ \etaV = \bigg(1-\frac{R_\textrm{E}}{|\nuV|}\bigg)\nuV.
   \]
   Sjølv om SIS er urealistisk, i og med at massen har uendeleg
   ustrekking, er modellen like fullt rekna som nyttig og mykje brukt i fysiske
   analysar.
\end{example}

Det er trivielt å simulera forvrengde bilete for ein gjeven kjelde- og linsemodell,
i alle fall dersom $\psi$ er deriverbar.
For kvar piksel $\nuV$ i det forvrengde biletet, kan me rekna ut posisjonen $\etaV$
og \emph{sampla} kjelda.
Utfordringa er å finna ein plausibel linsemodell $\hat\psi$ for eit observert forvrengt
bilete.
Det generelle tilfellet, der $\psi$ er ein vilkårleg funksjon, har uendeleg mange
fridomsgradar og er dermed ikkje mogleg å løysa.
Ei mogleg tilnærming er å føresetja ein konkret linsemodell (t.d.\ SIS)
der ein berre treng å fastsetja nokon få parameter (som einsteinradien).
\citet{hezaveh17} brukte t.d.\ maskinlæring for å finna parametra i ein
SIE-modell (singulær isotermisk ellipse).
I praksis står me derimot gjerne overfor sokalla klyngelinser,
der linsemassen er fordelt på fleire ulike objekt, kvar med sine parameter,
og dette krev ein meir samansett modell.

I dette arbeidet tek me ei alternativ tilnærming.
I staden for å fastsetja linsepotensialet $\psi$ over heile definisjonsområdet,
freistar me å skildra $\psi$ lokalt rundt eitt einskild punkt $\xiV=\nuV\cdot\DL/\DS$
i det forvrengte biletet.  Dersom me kan gjera dette for fleire galaksar som er påverka
av den same linsa, kan me i neste omgang freista å rekonstruera $\psi$ globalt
basert på fleire lokale skildringar.

\section{Roulettemodellen}\label{sec:roulette}

Rouletteformalismen vart introdusert av \citet{clarkson16a} og utdjupa av
\citet{Clarkson_2016_II}. 
I prinsippet er det ei taylorutviding av $\psi$ rundt eit punkt $\xiV$ 
i linseplanet, og
teknikken er godt etablert for svake linser der ein berre treng nokre få
taylorledd for å få ei rimeleg nøyaktig tilnærming.
Clarkson utvider denne teknikken for sterke linser ved å bruka fleire ledd,
i kontrast til \citet{fleury17} som utvider sterklinseteknikkar for
svake linser.

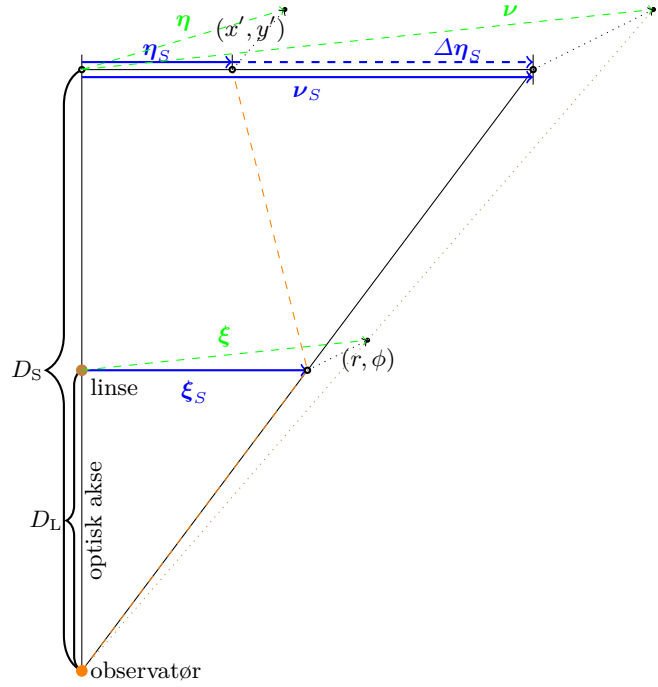
\begin{figure}[bt]
   \begin{center}
      \begin{tikzpicture}
   	\tikzset{point/.style={fill=white,draw=black,thick}}
	\tikzstyle{arrow}=[draw,->,blue,thick]
	\tikzstyle{lab}=[]
	\coordinate (origin) at (0,8);
   	\coordinate [label=right:observatør] (obs) at (0,0);
	\path [draw,black] (obs) -- (origin) ;

   	\coordinate [label=below right:linse] (lens) at (0,4);

	\coordinate (eta) at (2,8);
	\coordinate (eta1) at (2,8.1);
	\coordinate (nu) at (6,8);
	\coordinate (nu1) at (6,7.9);
	\coordinate (nu2) at (6,8.1);
	\coordinate (or1) at (0,7.9);
	\coordinate (or2) at (0,8.1);
	\path [draw,black] (obs) -- (nu) ;
	\path [draw,black] (origin) -- (nu) ;
	\path (lens) -- (obs) node [midway,rotate=90,below=-2pt] {optisk akse};

	\coordinate (xi) at (3,4);
	\path [draw,dashed,orange] (obs) -- (xi) -- (eta) ;
	\draw [arrow] (lens) -- (xi) node [lab,midway,below=0.1ex] {$\boldsymbol{\xi}_S$} ;


	\draw [decorate, decoration = {brace,amplitude=3.5ex}, thick] 
	      (obs) --  (origin) node [midway,left=0ex,xshift=-1.2em] {$D_\mathrm{S}$};
	\draw [decorate, decoration = {brace, amplitude=1.5ex}, thick] 
	      (obs) --  (lens) node [midway,left=0ex,xshift=-0.4em] {$D_\mathrm{L}$};

	\foreach \p in {eta,xi,nu,origin}
		\fill[point] (\p) circle (1pt);
	\foreach \p in {lens}
		\fill[draw=brown,fill=brown] (\p) circle (2pt);
	\foreach \p in {obs}
		\fill[draw=orange,fill=orange] (\p) circle (2pt);
        \draw [draw] (origin) -- +(90:0.2) ;
        \draw [draw] (eta) -- +(90:0.2) ;
        \draw [draw] (nu) -- +(90:0.2) ;
        \draw [draw] (nu) -- +(270:0.2) ;
	\draw [arrow] (or1) -> (nu1) node [lab,midway,below] {$\boldsymbol{\nu}_S$} ;
	\draw [arrow] (or2) -> (eta1) node [lab,midway,above=-3pt] {$\boldsymbol{\eta}_S$} ;
        \draw [arrow,dashed] (eta1) -> (nu2) node [lab,near end,above=-3pt] {$\Delta\boldsymbol{\eta}_S$} ;

	\coordinate  (etap) at (2.7,8.8) ;
	\coordinate  (nup) at (7.6,8.8) ;
	\coordinate  (xip) at (3.8,4.4) ;

	\foreach \p in {etap,nup,xip}
		\fill[draw=black,fill=black] (\p) circle (0.75pt);

	\draw [draw,dashed,green,->] (origin) -> (nup) node [near end,above] {$\boldsymbol{\nu}$} ;
	\draw [draw,dashed,green,->] (origin) -> (etap) node [midway,above] {$\boldsymbol{\eta}$} ;
	\draw [draw,dashed,green,->] (lens) -> (xip) node [midway,above] {$\boldsymbol{\xi}$} ;

	\draw [draw,dotted,->] (nu) -> (nup) ;
	\draw [draw,dotted,->] (xi) -> (xip) node [midway,black,right,xshift=-2pt,yshift=-2pt] {\footnotesize $(r,\phi)$} ;
	\draw [draw,dotted,->] (eta) -> (etap) node [midway,above,xshift=-3pt,yshift=-4pt] {\footnotesize $(x',y')$} ;
	\draw [draw,dotted,brown,thin] (obs) -> (nup) ;
\end{tikzpicture}
   \end{center}
   \caption{Geometrien i den flate himmelen med dei kritiske punkta i roulettemodellen.}
   \label{fig:model_2D}
\end{figure}

For å simulera forvrengde bilete i rouletteformalismen, bruker me fyrst
\emph{raytrace}-likninga for å finna punktet $\xiV_S$ i linseplanet,
som svarer til sentrum $\etaV_S$ i den opprinnelege kjelda.
I SIS-modellen er $\xiV_S$ gjeven som
\begin{align*}
   |\xiV_S| = 
   \frac{D_{\mathrm{L}}}{D_{\mathrm{S}}}\cdot|\nuV_S| =
   \frac{D_{\mathrm{L}}}{D_{\mathrm{S}}}\cdot|\etaV_S|
   + R_{\mathrm{E}}.
\end{align*}
Punktet $\xiV_S$ bruker me som referansepunkt for rouletteutvidinga
(jf.\ figur~\ref{fig:model_2D}).
Den fyrste rouletteamplityden $(\alpha_1^0,\beta_1^0)$ er då
avbøyinga $\nuV-\etaV$, 
\begin{align}
\label{Deta}
   \Delta\etaV
   & =\nuV-\etaV =-D_\textrm{S}\cdot(\alpha^0_1,\beta^0_1).
\end{align}
For å finna ljoset i eit vilkårleg punkt $\nuV$ nær $\nuV_S$, bruker me
roulettelikninga.
Me fylgjer Clarkson og
skriv punktet $\xiV$ i polarkoordinatar $(r,\phi)$ med origo i $\xiV_S$,
og $\etaV$ i kartesiske koordinatar $(x',y')$ med origo i $\etaV_S$.
Clarkson gjev oss då
\begin{align}
      \frac{\DL}{\DS}\cdot
   \begin{bmatrix} x' \\ y' \end{bmatrix} &=
   r\cdot\begin{bmatrix} \cos\phi \\ \sin\phi \end{bmatrix} 
    \label{eqn:general mapping}
      + \sum_{m=1}^{\infty} \frac{r^m}{m!\cdot\DL^{m-1}}
      \sum_{s=0}^{m+1} c_{m+s}
       \left(\alpha_s^m \boldsymbol{A}_{s} + \beta_s^m \boldsymbol{B}_{s} \right) 
       \begin{bmatrix} C^+ \\ C^- \end{bmatrix}
\end{align}
der
\begin{align}
   C^\pm &= \pm \frac{s}{m+1},\\
   c_{m+s} &= 
      \frac{1 - (-1)^{m+s}}{4} =
   \begin{cases}
      0, \quad m+s \text{ er jamn},\\
      \frac12, \quad m+s \text{ er odde},
   \end{cases}
\end{align}
og
\begin{align*}
    \boldsymbol{A}_{s} &= \begin{bmatrix}
    \cos{(s-1)\phi} & \cos{(s+1)\phi} \\ 
    -\sin{(s-1)\phi} &  \sin{(s+1)\phi} \end{bmatrix},
    &\quad
    \boldsymbol{B}_{s} &=
    \begin{bmatrix} 
        \sin{(s-1)\phi} & \sin{(s+1)\phi} \\
        \cos{(s-1)\phi} & -\cos{(s+1)\phi} 
    \end{bmatrix}.
\end{align*}
   Me kan leggja merke til at $\alpha_s^m$ og $\beta_s^m$ forsvinn når
   $m+s$ er jamn.
   I svak linsing er det normalt tilstrekkeleg å bruka det nullte 
   ($r(\cos\phi,\sin\phi)$) og fyrste ($m=1$) leddet i
   \eqref{eqn:general mapping}.
   Det nullte leddet gjev forstørringa medan det fyrste, kjend
   som skjæret (eller \emph{shear}), gjev elongeringa.
   Den kjende Kaiser-Squires-likninga fortel oss korleis ein kan 
   rekonstruera linsepotensialet $\psi$ frå desse to ledda
   \citep{normann2020}.

   Ledda av høgare orden ($m>1$) gjev informasjon om krumminga som oppstår
   i sterke linser.  Enkelt sagt vil ein rund galakse sjå oval ut bak
   ei svak linse.
   Bak ei sterk linse vert galaksen krum som ein banan\footnote{Eller kan henda meir som ei nyrebønne? Forfattarane er enno ikkje samde på dette punktet...}.
   Svært sterke linseeffektar gjev ein sokalla einsteinring, der den fjerne 
   galaksen ser ut som ein ring som strekker seg rundt linsa.
   Når Clarkson utvider den kjende svaklinsemodellen for sterke linser,
   så reiser det spørsmålet om ogso Kaiser-Squires-likninga
   kan generaliserast.

   Koeffisientane $\alpha_m^s$ og $\beta_m^s$ i likningane over kallar me 
   rouletteamplitydane.
   \citet{normann2020} har utleidd rekursive formlar for å finna algebraiske
   uttrykk for $(\alpha_m^s,\beta_m^s)$ som funksjonar av $\xi$ for alle $m$ og $s$.
   Implementasjonen vår, slik han vart presentert på ECMS 2023 \citep{hgs2023ecms},
   gjorde to ting.
   For det fyrste reknar han ut rouletteamplitydane for SIS, og for det andre
   simulerer han forvrengde bilete iht.\ likning \eqref{eqn:general mapping}.

   \begin{figure}[tb]
      \begin{subfigure}{0.24\textwidth}
         \includegraphics[width=\textwidth]{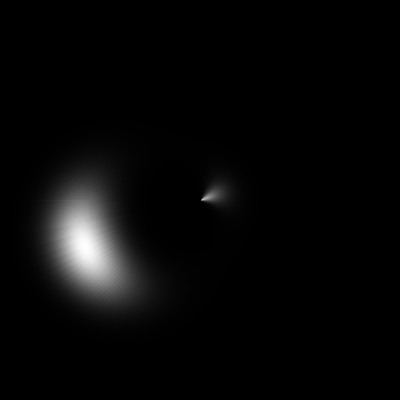}
         \caption{\emph{Raytrace}}
      \end{subfigure}
      \begin{subfigure}{0.24\textwidth}
         \includegraphics[width=\textwidth]{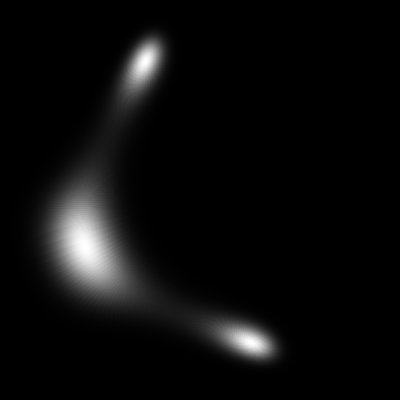}
         \caption{Roulette $m\le3$}
      \end{subfigure}
      \begin{subfigure}{0.24\textwidth}
         \includegraphics[width=\textwidth]{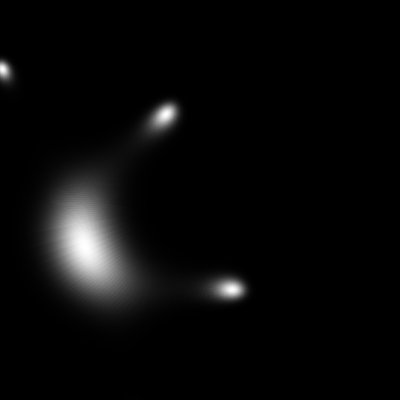}
         \caption{Roulette $m\le5$}
      \end{subfigure}
      \begin{subfigure}{0.24\textwidth}
         \includegraphics[width=\textwidth]{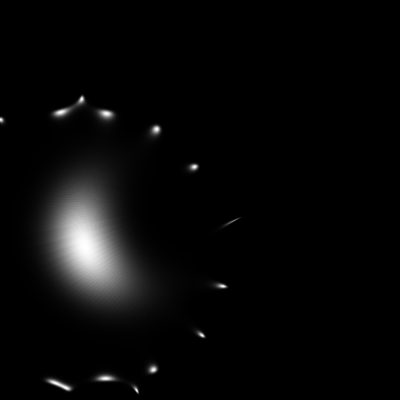}
         \caption{Roulette $m\le15$}
      \end{subfigure}
      \caption{Simulering i roulettemodellen og samanlikning med \emph{raytrace}.}
      \label{fig:roulette}
   \end{figure}

   Me kan samanlikna simulering i roulettemodellen med den eksakte 
   \emph{raytrace}-simuleringa i figur~\ref{fig:roulette}.
   Den uendelege rekkja i \eqref{eqn:general mapping} må trunkerast
   i praksis, so me har simulert for $m\le3$, $m\le5$ og $m\le15$.
   Småbileta som ligg i ein ring rundt hovudbiletet er numeriske artifaktar.
   Ein kan visa at der alltid er $m+1$ slike bilete når $m$ er odde og $m$
   når $m$ er jamn, og dei legg seg som ein ring rundt $\xiV_S$.  
   Radien i denne sokalla konvergensringen går mot $|\xiV_S|$ 
   når $m$ går mot uendeleg~\citep{Clarkson_2016_II}.
   Det er sjølvsagt mogleg å maskera ut modellartifaktane sidan me veit kvar
   dei ligg.

   Me kan merka oss at roulettemodellen ikkje får med bibiletet som ligg
   nær sentrum i \emph{raytrace}-biletet.  Bibiletet er eit resultat av
   ljos som går den lange vegen rundt linsa på motsett side.
   Dette bibiletet hamnar alltid utanfor konvergensringen i roulette\footnote{%
      Det er mogleg å bruka rouletteformalismen rundt eit anna punkt, 
      og dermed teikna bibiletet, men sidan konvergensringen alltid går
      gjennom linsa, vil ein aldri kunna få med båe bileta i same
      roulettemodell.}
   Bortsett fra bibiletet og dei falske bileta langs konvergensringen,
   ser med ei høveleg god tilnærming med fem ledd og med 15 ledd 
   er biletet nær perfekt.

   \begin{remark}
      Der finst ei rekkje simulatorar for gravitasjonslinser.
      Særleg Lenstronomy \citep{birrer18} og
      PyAutoLens \citep{Nightingale2021} er populære.
      Dei \emph{tutorials} som finst for PyAutoLens gjev god innsikt
      i korleis ein kan byggja opp samansette linsemodellar og 
      tilpasse parameter til empiriske bilete.
      Det som er nytt i tilnærminga vår er implementasjonen av
      rouletteformalismen og \emph{lokal} skildring av linsepotensialet
      ogso for sterke linser.  
   \end{remark}

\section{Rekonstruksjonsmodell}

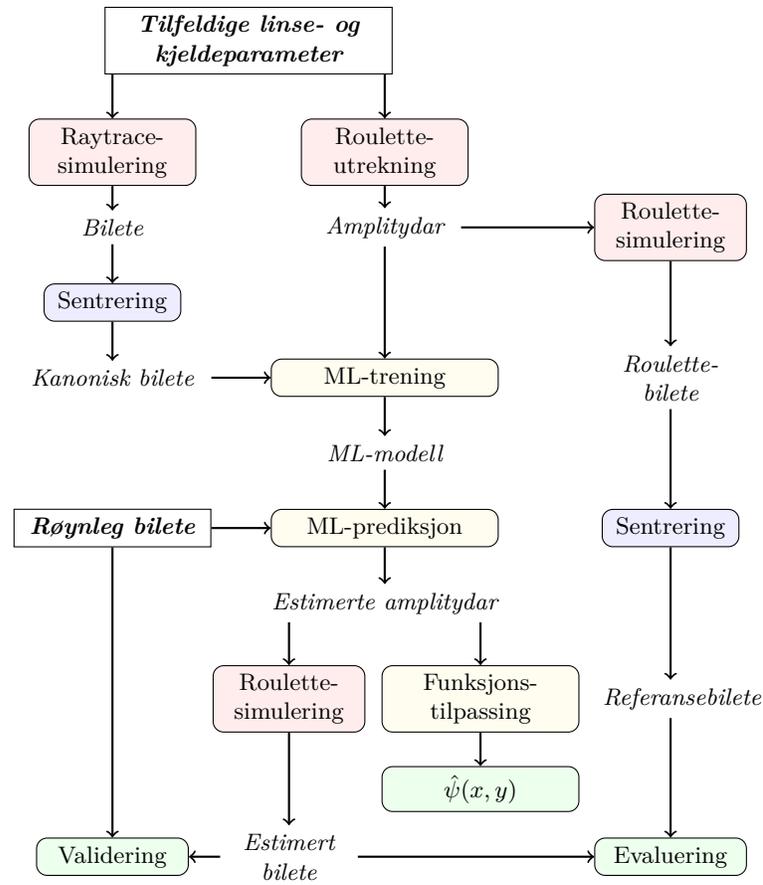
\begin{figure}
   \begin{center}
   \def\pb#1#2{\parbox{#1}{\centering #2}}
\begin{tikzpicture}
   \tikzstyle{val}=[draw,rounded corners,fill=green!7]
   \tikzstyle{op}=[draw,rounded corners,fill=blue!7]
   \tikzstyle{open}=[draw,rounded corners,fill=yellow!7]
   \tikzstyle{sw}=[draw,rounded corners,fill=red!7]
   \tikzstyle{dat}=[font=\itshape]
   \tikzstyle{input}=[draw,font=\bf\itshape]
   \tikzstyle{arrow}=[draw,thick,->]
   \node (rnd) [input] {\pb{36mm}{Tilfeldige linse- og kjeldeparameter}} ;
   \node (ray) [sw,below left of=rnd,node distance=21mm,xshift=-1em] {\pb{20mm}{Raytrace-simulering}} ;
   \node (rcalc) [sw,below right of=rnd,node distance=21mm,xshift=1em] {\pb{20mm}{Roulette-utrekning}} ;
   \node (amp) [dat,below of=rcalc] {\pb{18mm}{Amplitydar}} ;
   \node (img) [dat,below of=ray] {\pb{32mm}{Bilete}} ;
   \path [arrow] (rnd.south -| ray) -> (ray) ;
   \path [arrow] (rnd.south -| rcalc) -> (rcalc) ;
   \path [arrow] (ray) -> (img) ;
   \path [arrow] (rcalc) -> (amp) ;
   \node (c1) [op,below of=img] {\pb{16mm}{Sentrering}} ;
   \node (cimg) [dat,below of=c1] {\pb{24mm}{Kanonisk bilete}} ;
   \path [arrow] (img) -> (c1) ;
   \path [arrow] (c1) -> (cimg) ;
   \node (train) at (cimg -| amp) [open] {\pb{28mm}{ML-trening}} ;
   \node (model) [dat,below of=train] {\pb{28mm}{ML-modell}} ;
   \node (pred) [open,below of=model] {\pb{28mm}{ML-prediksjon}} ;
   \path [arrow] (amp) -> (train.north -| amp.south) ;
   \path [arrow] (cimg) -> (train) ;
   \path [arrow] (train) -> (model) ;
   \path [arrow] (model) -> (pred) ;
   \node (est) [dat,below of=pred] {\pb{34mm}{Estimerte amplitydar}} ;
   \path [arrow] (pred) -> (est) ;

   \node (fr) [open,below right of=est,node distance=18mm] {\pb{24mm}{Funksjons\-tilpassing}} ;

   \node (rsim) [sw,right of=amp,node distance=38mm] {\pb{18mm}{Roulette-simulering}} ;
   \node (rimg) at (train -| rsim) [dat] {\pb{18mm}{Roulette-bilete}} ;
   \path [arrow] (amp) -> (rsim) ;
   \path [arrow] (rsim) -> (rimg) ;
   \node (c2) at (pred -| rimg) [op] {\pb{16mm}{Sentrering}} ;
   \node (crimg) at (fr -| c2) [dat] {\pb{18mm}{Referansebilete}} ;
   \path [arrow] (rimg) -> (c2) ;
   \path [arrow] (c2) -> (crimg) ;

   \node (rsim2) [sw,below left of=est,node distance=18mm] {\pb{18mm}{Roulette-simulering}} ;
   \node (estim) [dat,below of=rsim2,node distance=21mm] {\pb{16mm}{Estimert bilete}} ;
   \path [arrow] (est.south -| rsim2.north) -- (rsim2) ;
   \path [arrow] (rsim2) -> (estim) ;

   \node (eval) at (estim -| crimg) [val] {\pb{18mm}{Evaluering}} ;
   \path [arrow] (estim) -> (eval) ;
   \path [arrow] (crimg.south -| eval) -> (eval) ;

   \node (real) at (cimg |- pred) [input] {\pb{24mm}{Røynleg bilete}} ;
   \node (val) at (estim -| real) [val] {\pb{18mm}{Validering}} ;
   \path [arrow] (estim) -> (val) ;
   \path [arrow] (real.south -| val) -> (val) ;
   \path [arrow] (real) -- (pred) ;

   \node (psi) [val,below of=fr,node distance=12mm] {\pb{24mm}{$\hat\psi(x,y)$}} ;
   \path [arrow] (est.south -| fr) -- (fr) ;
   \path [arrow] (fr) -- (psi) ;
\end{tikzpicture}
   \end{center}
   \caption{Prosessmodell.  Lokal rekonstruksjon av linsepotential.}
   \label{fig:sys}
\end{figure}

   Føresetnaden for å kunna bruka ei generalisering av Kaiser-Squires,
   er at me kan estimera rouletteamplitydane frå observerte bilete.
   Det er mogleg for skjæret, men for høgare orden finst der
   ingen analytisk metode i dag.  
   Simulatoren som me har drøfta over gjer det derimot mogleg å generera
   store datasett, og det er verd å sjå om maskinlæring kan estimera
   rouletteamplitydane.
   Figur~\ref{fig:sys} viser ein fullstendig prosess som kombinerer 
   rouletteformalismen med maskinlæring.

   Dei raude boksane bruker simulatoren \emph{CosmoSim} som me har
   implementert, testa og gjort tilgjengeleg som open kjeldekode\footnote{%
      Publisert på github.
      Drøftinga her tek utgangspunkt i versjon 2.3.0:
      \url{https://github.com/CosmoAI-AES/CosmoSim/releases/tag/v2.3.0}.}.
   For å generera treningsdata til maskinlæringa bruker me 
   \emph{raytrace}-likninga \eqref{eq:raytrace} for å generera
   bilete (\emph{input}) og roulette-utrekninga (sjå avsnitt~\ref{sec:roulette})
   for å generera amplitydane som er \emph{output} (\emph{ground truth})
   i maskinlæringa.
   Roulettesimulatoren gjer det mogleg å simulera forvrengde bilete direkte
   frå rouletteamplitydane, utan å ha nokon eksakt linsemodell.
   Dette gjev eit ekstra høve til å validera resultat, både under trening og
   i faktisk bruk.
   Bilete frå roulettesimuleringa gjev ei referansesanning som kan samanliknast
   med \emph{input} til maskinlæringa.  Dersom det estimerte biletet ikkje stemmer
   me det røynlege biletet, veit me at dei estimerte amplitydane er 
   unøyaktige eller feile.

   Me kjem tilbake til maskinlæringsdelen i avsnitt~\ref{sec:ml}.
   Det som me har kalt funksjonstilpassing i figuren, svarer til Kaiser-Squires-likninga.
   Basert på estimerte rouletteamplitydar, ynskjer me å finna eit estimat $\hat\psi$
   for linsepotensialet.
   Det er viktig å merka seg at eitt sett med rouletteamplitydar berre gjev lokal 
   informasjon om linsepotentialet rundt eitt punkt.
   Dersom me skal estimera $\psi$ over heile definisjonsområdet vil me måtte
   bruka bilete av fleire forvrengde galaksar og rouletteamplitydane 
   rundt kvar av dei.  Dette er ofte mogleg i empiriske billete, der ein kan sjå
   mange galaksar som er forvrengde av den same linsa.
   Denne funksjonstilpassinga står att som eit ope problem.

   Bileta vert sentrerte for å unngå å lekkja informasjon om linseposisjonen 
   relativt til den synlege galaksen.
   Me reknar ljossentrum i biletet ved å ta gjennomsnitt av pikselindeksane 
   vekta med ljosintensiteten.  Biletet vert so translatert slik at ljossentrum
   vert sentrum i biletet.
   Dette gjev ein kanonisk form som me ogso kan finna for røynlege bilete

   \subsection{Rekonstruksjonssimulatoren}

   Som me har nemnd bruker \emph{raytrace}-simulatoren eit koordinatsystem
   med origo på den optiske aksen (gjennom linsa).
   Når me simulerer med utgangspunkt i rouletteamplitydane er linsesentrum
   i prinsippet ukjend.  Rett nok kan der vera noko ljos som gjev ein 
   omtrentleg linseposisjon, men me vil ikkje gå ut frå at dette gjev
   ein tilstrekkeleg presis posisjon.
   Det er òg grunnen til at biletet vert sentrert rundt ljossentrum.
   
   For å rekonstruera det forvrengde biletet treng me informasjon
   om kjelda, i tillegg til rouletteamplitydane som gjev informasjon 
   om linsa.
   Kjeldesentrum er representert som $\xiV'=(\xi_1',\xi_2')$ relativt
   til origo i ljossentrum.
   For ei sfærisk kjelde treng me dessutan storleiken, som i \emph{CosmoSim}
   er representert som standardavviket $\sigma$ i ein gaussisk ljosfordeling.
   Andre kjeldemodellar krev fleire parameter, men det har me ikkje testa
   i denne studien.

   Dei tre parametra $\sigma$, $\xi_1'$ og $\xi_2'$ vert inkludert i 
   datasettet i maskinlæringa, saman med rouletteamplitydane.

   \subsection{Oversikt over implementasjonen}

   Sjølve simulatoren i \emph{CosmoSim} er eit bibliotek i C++.
   To brukargrensesnitt er implementerte i Python.
   Kommandolinegrensesnittet som me bruker her, er designa for
   satsgenerering av bilete.  GUI-verktyet drøfta me på
   ECMS \citep{hgs2023ecms}, og det er ikkje relevant her.

   Simulatoren fylgjer ein enkel objektorientert struktur.
   Eit \emph{Source}-objekt definerer den synlege galaksen
   og genererer biletet slik det hadde sett ut utan gravitasjonslinsa.
   Her har me implementert underklasser for sfæriske og elliptiske linser,
   samt ein trefarga trekant til illustrasjonsformål.

   Rouletteamplitydane vert rekna ut symbolsk, vha.\ sympy-biblioteket
   i python, og skrive til ei fil som vert lese i C++-biblioteket,
   som so evaluerer amplitydane i konkrete punkt.

   Eit \emph{Lens}-objekt definererer linsemodellen med alle dei analytiske
   eigenskapane som er kjende.  Særleg er det $\psi$ med dei to fyrsteordens
   partiellderiverte og rouletteamplitydane som trengst i simulatoren.
   Her har me førebels berre implementert ei underklasse for SIS,
   samt ei klasse som \emph{samplar} ein anna linsemodell og reknar med
   numerisk derivasjon.  Sistnemnde kan vera nyttig til meir kompliserte
   linsemodellar der symbolsk derivasjon ikkje er mogleg innanfor rimeleg tid.

   Sjølve simulatoren er ei underklasse av \emph{LensModell}.
   Her har me implementert \emph{RaytraceModel} og \emph{RouletteModel}.
   Dei tek ei \emph{Source} og ei \emph{Lens} og bruker hhv.\
   \eqref{eq:raytrace} og \eqref{eqn:general mapping} for å transformera
   biletet frå \emph{Source} til eit realistisk forvrengd bilete slik
   det vert observert.
   \emph{RaytraceModel} hentar dei partiellderiverte av $\psi$
   frå \emph{Lens}-objektet medan \emph{RouletteModel} bruker
   rouletteamplitydane.

   For å rekonstruera biletet frå rouletteamplitydane aleine, bruker
   me ei tredje \emph{LensModel}-klasse,
   \emph{RouletteRegenerator}, som ikkje bruker noko \emph{Lens}-objekt.
   Simuleringa er den same som i \emph{RouletteModel}, men rouletteamplitydane
   vert sette direkte i \emph{RouletteRegenerator}, ingen annan informasjon
   om linsa vert tilgjengeleg.

   Kommandolineprogrammet, slik me normalt bruker det, tek ein CSV-fil
   med linse- og kjeldeparameter og genererer eitt bilete per rad.
   I tillegg kan det, i same prosess, generera ei ny CSV-fil med
   rouletteamplitydane.

\section{Resultat}
\label{sec:ml}

For å demonstrera at rammeverket har noko for seg, presenterer me ein
konkret serie av testar.  
Me har førebels gjort lite for å optimalisera maskinlæringsoppsettet 
eller utfallsrommet for datasettet, og ein lyt difor lesa det som eit 
døme og ikkje som eit råd til endeleg løysing.

\subsection{Biletgenerering}

   Datasettet simulerer ei sfærisk kjelde sett gjennom ei SIS-linse.
   Relativ avstand til linsa, $\chi=\DL/\DS$ set me konstant lik $0{,}5$.
   Dette gjev fire variable parameter:
   storleiken (standardavviket) $\sigma$ for kjelda, einsteinradien $R_E$,
   og kjeldeposisjonen som me skriv i polarkoordinatar $(R,\phi)$.  Desse
   dreg me uniformt tilfeldig frå fylgjande sannsynsfordeling:
   \begin{align*}
      \sigma &\in\{1,2,\ldots60\}, &\quad
      R_E    &\in \{5,6,\ldots 50\}, \\
      \phi   &\in \{ 0^\circ,1^\circ,\ldots359^\circ \}&\quad
      R      &\in \{R_E,R_E+1,\ldots100\}
   \end{align*}
   Polarkoordinatane vert rekna om til kartesiske koordinatar $(x,y)$ som 
   vert brukte i maskinlæringa, men me merker oss at det er avstanden $R$
   til origo som er uniformt fordelt, ikkje $x$- og $y$-koordinatane.

   For å vurdera køyretid, har me køyrd sju samtidige satsar
   à $20\,000$ bilete.  Kvar sats tek 59--65 minutt
   sanntid og 138--170 minutt CPU-tid på åtte kjernar\footnote{%
      Me har brukt tungreiningsklynga IDUN ved NTNU, og me har ikkje
      registrert kva node som har vore tildelt og dermed kjenner me
      ikkje prosessorspesifikasjonane.}.
   Dette er overkommeleg, og langt fleire bilete enn me bruker i den
   vidare testen.
   Forsøk med fleire samtidige jobbar gjev lengre køyretid, og det er
   rimeleg å gå ut frå at flaskehalsen er skriving til disk.

   Skriptet \texttt{datagen.py} genererer både bileta og CSV-filen
   med rouletteamplitydar; som fylgjar
   \begin{verbatim}
   python3 CosmoSimPy/datagen.py -D <biletkatalog> \
      -C -Z 800 -z 400 --lensmode SIS --modelmode Raytrace \
      --nterms 5 --outfile roulette.csv --csvfile dataset.csv \
      --xireference
   \end{verbatim}
   Her lagar me bilete på $800\times800$ som vert klipt til $400\times400$ etter
   sentrering.  Rouletteamplitydane vert funne for $m\le5$.
   Opsjonen \texttt{--xireference} seier at rouletteamplitydane vert rekna
   i den tilsynelatende posisjonen til sentrum i kjelda.  Dette ligg normalt
   ikkje i ljossentrum av det forvrengde biletet.

   Me simulerer dei forvrengde bileta i rouletteformalismen som referanse.
   Fordi denne simulatoren har eit anna API, vert dette gjort med eit
   eige skript, som fylgjer.
   \begin{verbatim}
   python3 CosmoSimPy/roulettegen.py -D <roulettebiletekatalog> \
           -n 5 -Z 400 --csvfile roulette.csv --xireference
   \end{verbatim}
   Køyretida er 29--30 min.\ sanntid og 98--99 min.\ CPU-tid per sats
   på $20\,000$ bilete.

   Datasettet omfattar 33 søyler som trengst for å kunne resimulera
   det forvrengde biletet i rouletteformalismen.
   Der er 30 rouletteamplitydar for $m=0,\ldots,5$, samt
   linseparameteren $\sigma$ og 
   kjeldesentrum $\xiV'$
   relativt sentrum i biletet (ljossentrum).

\subsection{Maskinlæring}

   Til maskinlæringa har me brukt Inception v3, modifisert for å ta
   éin kanal (gråtone) inn og gje regresjonsdata ut, i staden for
   klassifisering.  Modifikasjonane er tekne frå arbeidet åt
   \citet{cosmoai2022bsc}.
   Optimeringsalgoritma er Adams der alle parameter har initialinstillingar
   bortsett frå læringsraten $\alpha=0{,}0001$.
   Som tapsfunksjon bruker me gjennomsnittleg kvadratfeil (MSE).
   Me har brukt 4000 bilete til trening og $10\,000$ til testing.
   Med 50 epokar tek dette under to timar på ein NVIDIA A100 (GPU).

\begin{figure}
   \begin{subfigure}{0.5\textwidth}
   \begin{center}
   \includegraphics[width=\textwidth]{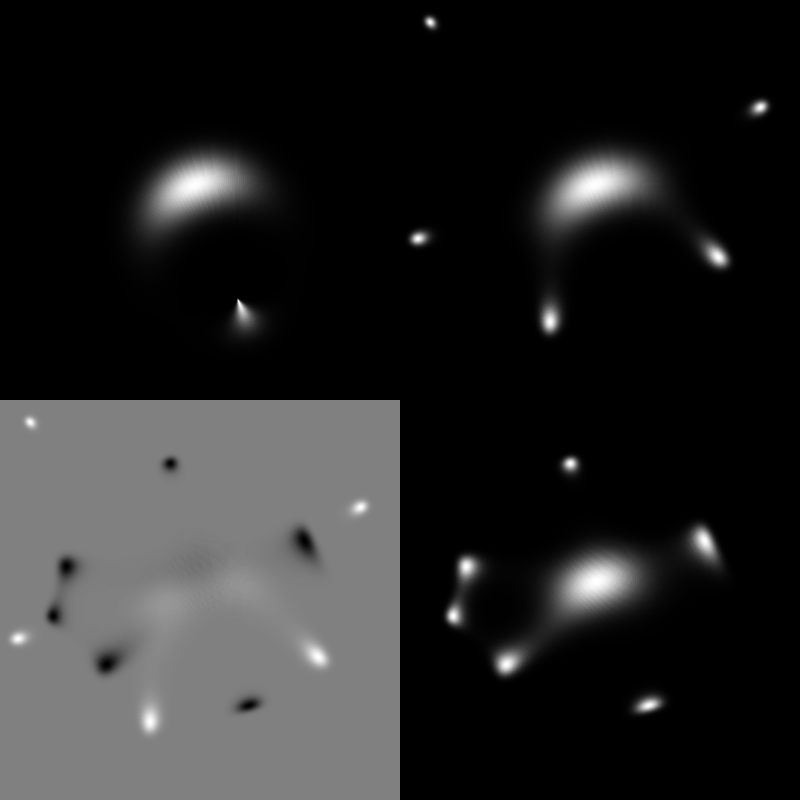}
   \end{center}
   \caption{Døme på rimeleg biletestimat.}
   \label{fig:good}
   \end{subfigure}
   \begin{subfigure}{0.5\textwidth}
   \begin{center}
   \includegraphics[width=\textwidth]{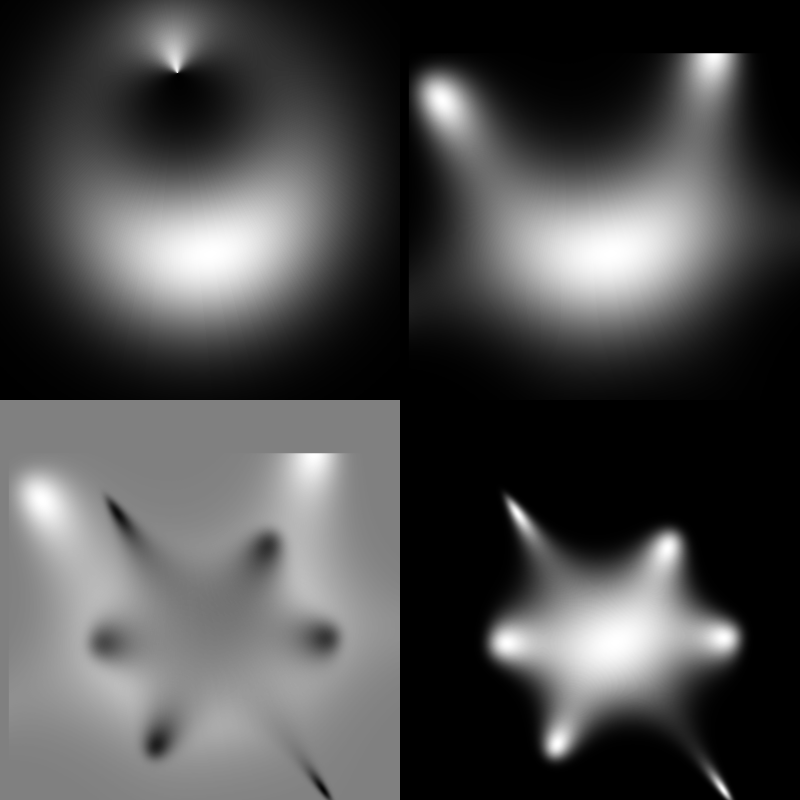}
   \end{center}
   \caption{Døme på dårleg estimert bilete.}
   \label{fig:bad}
   \end{subfigure}
   \caption{Samanlikning av rekonstruerte bilete.  I kvar montage har me
     opprinneleg bilete øvst til venstre.  Roulettesimulering frå faktiske
     amplitydar øvst til høgre og frå estimerte amplitydar nedst til høgre.
     Nedst til venstre ser me differansen mellom dei to roulettesimuleringane.}
   \label{fig:reconstruction}
\end{figure}

   Figur~\ref{fig:reconstruction} viser rekonstruerte bilete basert på
   estimerte amplitydar.  I det eine biletet, der kjelda er stor, ser me
   ingen gjenkjennelege drag.  I det andre, med ei mindre kjelde, ser me
   tydeleg at både retning frå origo og krumming er riktig, sjølv om
   rekonstruksjonen er langt frå nøyaktig.
   Me kan dermed slutta at det er mogleg å dra relevant informasjon ut
   av biletet vha.\ maskinlæring, og det er sannsynleg at ein kan finna
   betre resultat om ein legg arbeide i det.

   Ein kan stussa på at det rekonstruerte biletet i figur~\ref{fig:good}
   ikkje har dei falske bileta jamnt fordelt rundt konvergensringen.
   Dette skuldast sannsynlegvis at dei estimerte amplitydane ikkje treng
   å svara til ein kontinuerleg og deriverbar funksjon $\psi$.
   Estimeringsfeilen kan gje artifaktar som ikkje er moglege i utgangspunktet.

\begin{figure}
   \begin{subfigure}{0.5\textwidth}
      \begin{center}
      \includegraphics[width=\textwidth]{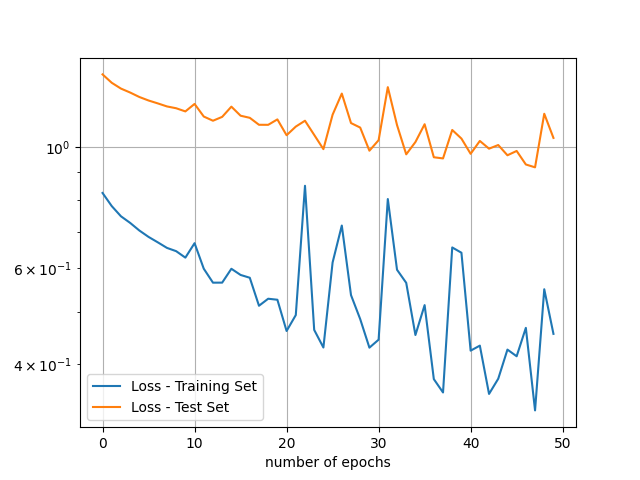}
      \end{center}
      \caption{\raggedright
      Tapsfunksjonen under trening for kvar epoke.}
   \end{subfigure}
   \hfill
   \begin{subfigure}{0.5\textwidth}
      \begin{center}
      \includegraphics[width=\textwidth]{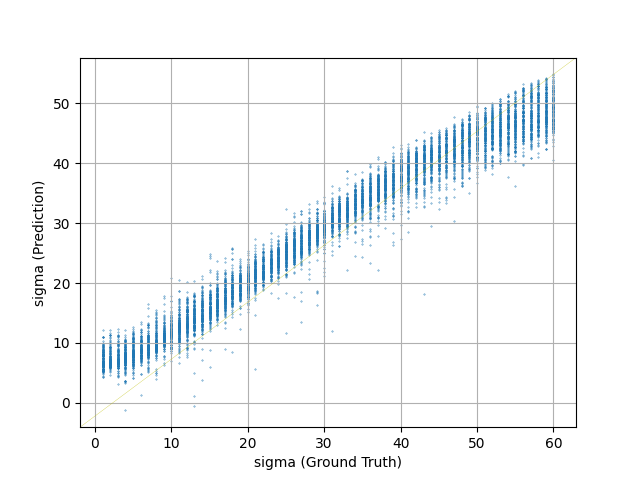}
      \end{center}
      \caption{\raggedright
      Samanlikning av estimat og \emph{ground truth} for $\sigma$.}
   \end{subfigure}
   \begin{subfigure}{0.5\textwidth}
      \begin{center}
      \includegraphics[width=\textwidth]{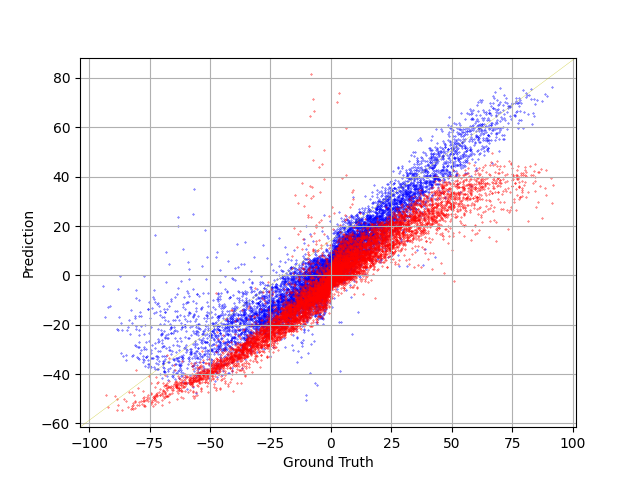}
      \end{center}
      \caption{\raggedright
      Samanlikning av estimat og \emph{ground truth} for $\xi'_1$
         (blått) og $\xi'_2$ (raudt).}
   \end{subfigure}
   \hfill
   \begin{subfigure}{0.5\textwidth}
     \begin{center}
     \includegraphics[width=\textwidth]{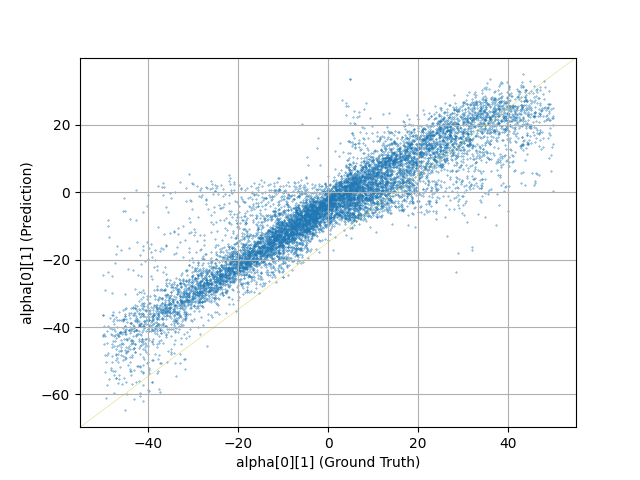}
     \end{center}
      \caption{\raggedright
      Samanlikning av estimat og \emph{ground truth} for $\alpha_1^0$.}
   \end{subfigure}
   \begin{subfigure}{0.5\textwidth}
     \begin{center}
     \includegraphics[width=\textwidth]{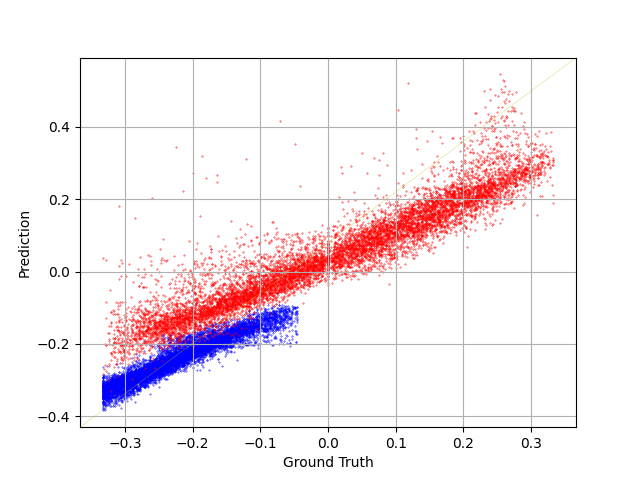}
     \end{center}
      \caption{\raggedright
      Samanlikning av estimat og \emph{ground truth} for $\alpha_0^1$
      (blått) og $\alpha_2^1$ (raudt).}
   \end{subfigure}
   \hfill
   \begin{subfigure}{0.5\textwidth}
     \begin{center}
     \includegraphics[width=\textwidth]{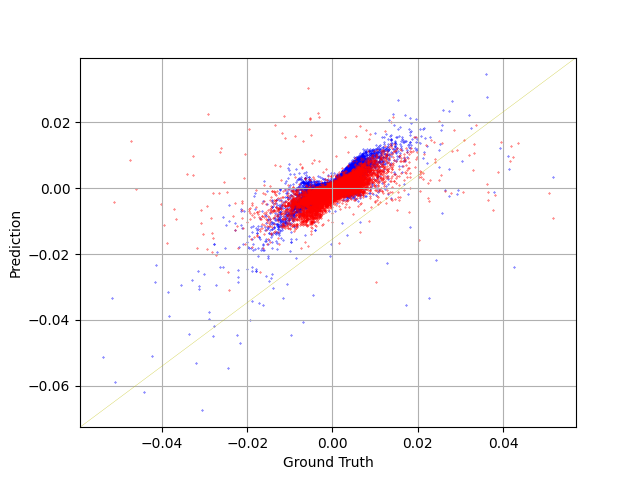}
     \end{center}
      \caption{\raggedright
      Samanlikning av estimat og \emph{ground truth} for $\alpha_1^2$
      (blått) og $\alpha_3^2$ (raudt).}
   \end{subfigure}
   \begin{subfigure}{0\textwidth}
   \end{subfigure}
   \caption{Evaluering av maskinlæringstesten.}
   \label{fig:eval}
\end{figure}

   Figur~\ref{fig:eval} viser kvantitativ evaluering av maskinlæringsoppsettet.
   Der er to ting som me skal merka oss.  
   For det fyrste vert $\xiV'$ systematisk underestimert, noko som òg gjev
   forskyvinga i det rekonstruerte bilete i figur~\ref{fig:good}.
   For det andre har me jamn betring i tapsfunksjonen i 
   ni epokar, før tapsfunksjonen tek til å svinga.
   Det tyder på at læringsraten er for høg etter dei ni epokane.

\section{Vegen vidare}

   Me har etablert eit rammeverk for å arbeida med rekonstruksjon
   av rouletteamplitydar vha.\ maskinlæring, 
   Dette opnar ei lang rekkje gode problem for vidare forsking.
   Innanfor maskinlæring gjenstår arbeidet med å finna god
   nettverksdesign og optimalisera hyperparameter.

   Me har framleis ikkje sett på testing med empiriske datasett.
   Dette er krevjande fordi astronomiske avstandar gjev låg oppløysing
   og ein må ta omsyn til optiske artifaktar i teleskopa.
   Tilretteleggjing av datasett frå røynda og tilpassing av treningssettet
   til empiriske data 
   er den største og kanskje mest spanande utfordringa framover.

   Innanfor matematisk fysikk gjenstår arbeidet med å generalisera
   Kaiser-Squires eller utarbeida andre teknikkar for å rekonstruera 
   linsepotentialet.
   Til sist vil det òg vera nyttig å utvida simulatoren med andre
   linse- og kjeldemodellar.

\printbibliography

\end{document}